\documentstyle[pra,aps,preprint]{revtex}
\begin{document}
\draft
\title{An Atom Laser Based on Raman
Transitions}
\author{G.M. Moy \cite{email:Moy}, J.J. Hope and C.M.
Savage}
\address{Department of Physics and Theoretical Physics,
The
Australian National University, \\
Australian Capital Territory 0200,
Australia.}
\date{\today}
\maketitle

\begin{abstract}
In this paper we present an atom laser scheme using a Raman transition 
for the output coupling of atoms.  A beam of thermal atoms (bosons) in 
a metastable atomic state is pumped into a multimode atomic cavity.  
This cavity is coupled through spontaneous emission to another cavity 
for the atomic ground state.  Above a certain threshold pumping rate a 
large number of atoms build up in the lowest energy state of the 
second cavity, while the higher energy states remain unpopulated.  
Atoms are then coupled to the outside of the cavity with a Raman 
transition.  This changes the internal level of the atom and imparts a 
momentum kick, allowing the atoms to leave the system.  We propose an 
implementation of our scheme using hollow optical fiber atom 
waveguides.
\end{abstract}


\narrowtext

\section{Introduction} \label{sec:Intro}

There has been much recent interest in the design of a device that 
would emit a coherent beam of bosonic atoms - an atom laser.  An atom 
laser would have many applications in atom optics including atom 
lithography and nanofabrication, as well as fundamental tests of 
quantum mechanics such as those involving atom interferometry.  A 
number of theoretical atom laser schemes have already been proposed 
\cite{Holland95,Wiseman95a,Wiseman95b,Guzman96,Spreeuw95,Olshanii95}.  
These have involved some method of cooling atoms in an atomic cavity, 
and a schematic model for coupling the atoms to the external freely 
traveling atomic modes.  Our approach differs from previous work since 
we model a particular atom laser scheme emphasizing the importance of 
output coupling.  Our model is quantitative and based on rate 
equations.  There is a lasing threshold analogous to that of the 
optical laser.  We consider a method for output coupling based on 
changing the internal atomic state of the atoms using a Raman 
transition in a spatially localized region.

We propose an implementation of our atom laser scheme using hollow 
optical fibers.  This has several advantages, including providing a 
directed output beam, and minimization of the reabsorption of 
spontaneously emitted photons.

In recent experiments which have produced a Bose-Einstein condensate 
\cite{Anderson95,Bradley95,Davis95,Mewes96}, a macroscopic number of 
atoms have been condensed in the ground state of an atomic trap.  This 
has generated increased interest in atom lasers, which also involve 
the production of large occupation numbers in a single quantum 
mechanical state.  In the reported theoretical atom laser models the 
ground state of an atomic trap or cavity (the lasing mode) is filled 
with a large number of atoms by using the higher energy modes of the 
trap as a continuously pumped atomic source.  The coupling between 
higher energy modes and the laser mode is achieved through cooling.  
In reference \cite{Wiseman95a} the mechanism is dark state cooling, 
and atoms are transferred from the source to the lasing mode 
irreversibly by spontaneous emission.  Spreeuw {\it et al.} 
\cite{Spreeuw95} use precooled atoms and a pump followed by 
spontaneous emission to transfer atoms into the lasing mode.  Holland 
{\it et al.} \cite{Holland95} and Guzman {\it et al.} \cite{Guzman96} 
use inelastic binary collisions to transfer atoms from the source mode 
to the lasing mode.  In reference \cite{Holland95}, two atoms collide 
to produce one atom in the lasing mode, and another in a higher energy 
mode.  This process is made irreversible by using evaporative cooling 
to rapidly remove the higher energy atom from the system.  A 
conceptually similar method is used in \cite{Guzman96}.  The number of 
atoms in the lasing mode depends on the pumping and loss rates.  Above 
threshold the number of atoms in the lasing mode saturates.  This 
produces a reasonably well defined number of atoms in a single cavity 
mode.

The atom laser proposals mentioned above have been schematic in nature 
and the only proposed methods of output coupling have involved either 
quantum mechanical tunneling\cite{Wiseman95a} or periodically turning 
off the cavity mirrors \cite{Guzman96}.  Wiseman {\it et al.} 
\cite{Wiseman95a} find that any physically realizable optical 
potential barrier confining the atoms leads to an extremely small 
tunneling rate.  Turning off the cavity mirrors, while effective for 
output coupling, will not provide a continuous beam.  We present an 
atom laser scheme using two atomic cavities - one for the source atoms 
in atomic level $|1 \rangle$ and a second for the lasing atoms in 
level $|2 \rangle$.  This second cavity has only one significantly 
populated mode - the ground state mode.  Raman transitions are used to 
change the state of the atoms in this mode to a non-trapped state, to 
allow the output coupling of the atoms from the lasing mode.

In Sec.  \ref{sec:Scheme} we give an overview of our basic atom laser 
scheme.  In Sec.  \ref{sec:Input} we investigate the input of bosons 
into the lasing mode.  We discuss the rate of transfer of the atoms 
from the pump cavity to the ground state of the lasing cavity, 
considering both the atomic transition rate due to the spontaneous 
emission of photons and the overlap between the wavefunctions of the 
two cavities.  In Sec.  \ref{sec:Output} we discuss our output 
coupling method.  This method could in principle be used for other 
atom laser schemes.  In Sec.  \ref{sec:Equations} we derive rate 
equations and compare our atom laser model with the standard optical 
laser and similar rate equations for the atom laser described by 
Spreeuw {\it et al.} \cite{Spreeuw95}.  We also investigate the 
threshold condition.  In Sec.  \ref{sec:Implementation} we present an 
outline of a possible implementation of our basic scheme using hollow 
optical fibers.  Finally, in Sec.  \ref{sec:Conclusion} we summarize 
our conclusions and discuss some of the limitations of our work.

\section{Atom Laser Scheme} \label{sec:Scheme}

In this section we present an overview of our atom laser scheme.  The 
details are discussed in the following sections.  
The model consists of atoms with four energy levels, as outlined in 
Fig.  \ref{EnergyLevels}.  Level $|1 \rangle$ is the input pump level, 
level $|2 \rangle$ is the lasing level and level $|4 \rangle$ is the 
output level.  Level $|3 \rangle$ mediates the output coupling Raman 
transition.  There are two atomic cavities for confining atoms in 
states $|1 \rangle$ and $|2 \rangle$.  One of these cavities, the 
lasing cavity, traps atoms in the level $|2 \rangle$.  Only a single 
mode of this cavity becomes populated as we show in Sec.  
\ref{sec:Equations}.  We wish to build up a large number of atoms in 
this ground state mode, in an analogous way to the standard optical 
cavity in a laser.  The other cavity, the pump cavity, traps a large 
number of atoms in the internal metastable level $|1 \rangle$.  The 
two cavities are spatially overlapping, see Fig.  \ref{fiber}.

Initially the atoms are prepared in level $|1 \rangle$, cooled and 
injected into the pump cavity.  Injection could be achieved by a 
number of methods.  For example, a partially reflecting atomic cavity 
mirror could be employed.  Such atomic mirrors can be produced using 
the repulsive potentials created by blue detuned laser 
beams\cite{AtomMirror}.  The transmittivity of such a mirror may be 
very small for a practical cavity \cite{Wiseman95a}, however extremely 
large input fluxes of atoms are possible allowing useful numbers of 
atoms into the cavity.  Another possible input mechanism would be to 
inject the atoms into the cavity in a non-trapped atomic state and 
then employ a change of atomic state (for example using a Raman 
transition) to the trapped state.  This reverses our output coupling 
method.

Atoms change from the pump level $|1 \rangle$ to the lasing level, $|2 
\rangle$ at a rate $r_{12}$ in the absence of Bose-enhancement.  Atoms 
which make this change of level, however, do not necessarily become 
trapped in the ground state of the lasing cavity - the lasing mode.  
However the wavefunction overlap with this ground state is larger than 
the overlap with other higher energy states.  Tunneling losses out of 
this state are also lower.  Atoms that do transfer to the ground state 
of the lasing cavity are trapped, and so further transitions to this 
state will be enhanced by a factor of ($N_{21}+1$) where $N_{21}$ is 
the number of atoms in atomic level $|2 \rangle$ and the ground state 
of the cavity.  With suitable pumping of atoms into the system, a 
large number of atoms will build up in this single quantum mechanical 
state of the combined atomic and cavity system.  For the parameters 
considered in this paper (Sec.  \ref{sec:Implementation}) other higher 
energy modes of the cavity state are not significantly populated.

A Raman transition couples atoms out of the system.  Two lasers 
transfer the atoms from level $|2 \rangle$ to a final atomic level $|4 
\rangle$.  The lasers are confined to the lasing cavity, and are shone 
diagonally across the cavity so that they are counter-propagating in 
one direction (the longitudinal direction) and co-propagating in the 
transverse direction, Fig.  \ref{fiber}.  Thus a momentum kick is 
imparted on atoms in the longitudinal direction, but no net momentum 
kick occurs in the transverse direction.  The longitudinal momentum 
kick pushes the atoms out of the interaction region.  If the rate at 
which atoms leave the system due to this kick is much larger than the 
Raman transfer rate then we have an effective irreversible transfer of 
atoms from the lasing mode out of the system.

In Sec.  \ref{sec:Implementation} we present an implementation of the 
scheme using a hollow optical fiber to create the transverse 
confinement for the atoms, and focussed laser light to form the 
longitudinal atomic mirrors.  For a fiber with a hole of diameter 
$\approx 2 \mu \mbox{m}$ it is appropriate to model the confining 
potential in the transverse direction as a harmonic oscillator 
potential.  In the longitudinal direction we model the confining 
lasers for the pump cavity as a square well, and those of the lasing 
cavity as a harmonic oscillator.  One possible problem with coupling 
the pump and lasing cavity by spontaneous emission is photon 
reabsorption.  A simple argument suggests that reabsorption might 
reduce the number of atoms in the ground trap state.  However, Cirac 
and Lewenstein \cite{Cirac96} have shown that under certain conditions 
reabsorption may increase the number of atoms in the ground state.  
Moreover, reabsorption can be circumvented by ensuring that the lasing 
cavity is smaller than the mean free path of a photon, as suggested by 
Spreeuw {\it et al}.  \cite{Spreeuw95}.  Hollow optical fibers with 
hole diameters comparable to the photon wavelength achieve this.

\section{Population of the lasing mode} \label{sec:Input}

Atoms enter the system from a cooled thermal source.  The initial 
cavity is produced using atomic mirrors for atoms in the metastable 
level $|1 \rangle$.  We consider an initial rate of atoms entering 
this cavity which we call the pump rate,$r_{1}$.  The coupling between 
the pump and laser cavity occurs through spontaneous emission.  The 
rate of transfer of atoms from the pump cavity to the lasing cavity 
depends both on this atomic transition rate, $r_{12}$, and on the 
average wavefunction overlap between the modes of the pump cavity and 
the lasing cavity.  The average overlap between the pump cavity and 
the $j$th mode of the lasing cavity, $g_{j}$, is equal to the sum over 
all the pump cavity states of the probability of finding an atom in 
that state, multiplied by the overlap between it and the $j$th energy 
state of the lasing cavity,
\begin{equation}
g_{j} = \sum_{i=0}^{\infty} n_{i} g_{ij} / N_{1},
\label{Over}
\end{equation}
where $N_{1}$ is the number of atoms in level $|1 \rangle$ in any pump 
cavity state.  $g_{ij}$ is the value of the wavefunction overlap 
between the $i$th energy state of the pump cavity and the $j$th energy 
state of the lasing cavity, as outlined in the appendix.  This overlap 
includes the effect of the random momentum kick from the spontaneously 
emitted photon.  $n_{i}$ is the number of atoms in the $i$th pump 
cavity state.  The exact nature of the pump cavity distribution does 
not affect the qualitative behaviour of the atom laser scheme, so for 
calculational purposes we will assume that the population of states is 
a Bose-Einstein distribution,
\begin{equation}
n_{i} = \frac{z \exp
\left[-E_{i}/k_{b}T \right]}{1 - z 
\exp \left[-E_{i}/k_{b}T \right]},
\label{BE}
\end{equation}
where $E_{i}$ is the energy of the $i\mbox{th}$ cavity state, $k_{b}$ 
is the Boltzmann constant and $T$ the temperature.  $z$ is the 
fugacity, which is related to the chemical potential, and can be found 
by solving the equation $\sum_{i = 0}^{\infty} n_{i} = N_{1}$.  The 
total rate $R_{12j}$ at which atoms transfer from the pump cavity to 
the $j$th mode of the lasing cavity is given by
\begin{equation}
R_{12j} =
(N_{2j}+1) N_{1} ~ r_{12} ~ g_{j}, \label{TotRate}
\end{equation}
where
$(N_{2j}+1)$ is the Bose-enhancement factor due to the presence of
$N_{2j}$ bosons in level $|2 \rangle$ and the $j$th state of the lasing
cavity.  $g_{j}$ is given by Eq.  (\ref{Over}).

\section{Output
Coupling} \label{sec:Output}

Other schemes for atom lasers 
\cite{Holland95,Wiseman95a,Wiseman95b,Guzman96,Spreeuw95} have not 
given a description of a practical output coupling mechanism.  We 
present here a method of switching the atomic state of the atoms, 
using a Raman transition, to allow output coupling from the system.  A 
pair of lasers at frequencies $\omega_{1}$ and $\omega_{2}$ induce a 
Raman transition from level $|2\rangle$ to the output level $|4 
\rangle$ of the atom, see Fig.  \ref{EnergyLevels}.  This output level 
is not trapped.
 
The lasers are oriented as discussed in Sec.  \ref{sec:Scheme} so as 
to impart a net momentum kick to the atoms of magnitude $2\hbar k_{x}$ 
in a particular direction, $x$.  For the hollow fiber situation 
discussed in Sec.  \ref{sec:Implementation}, the $x$ direction is the 
longitudinal fiber direction, so that the atoms are pushed out of the 
fiber.  The lasers are on resonance with the Raman $|2 \rangle 
\leftrightarrow |4\rangle$ transition.  They are far detuned from the 
$|2\rangle \rightarrow |3 \rangle$ and $|4 \rangle \rightarrow |3 
\rangle$ resonances to reduce excitation to level $|3 \rangle$.  The 
Raman transition from level $|2 \rangle$ to level $|4 \rangle$ has a 
transition rate that depends on the Rabi frequencies, $\Omega_{1}$ and 
$\Omega_{2}$, of the two lasers for their respective transitions $|2 
\rangle \rightarrow |3 \rangle$ and $|4 \rangle \rightarrow |3 
\rangle$.  This rate also depends on the detuning of the two lasers 
from level $|3\rangle$.  This transition rate can be found from 
scattering theory \cite{scattering}
\begin{equation}
r_{24} =
\frac{\Omega_{1}^{2} \Omega_{2}^{2}}{8 \Delta^{2} 
 \left(1/t_{0} \right)
}, \label{r24}
\end{equation}
where $\Delta = \omega_{23}-\omega_{1}$ is the detuning and $t_{0}$ is 
the time scale on which atoms are irreversibly lost from the system 
due to the momentum kick.  $1/t_{0}$ is then the single atom loss rate 
from the system.  The atoms that are lost from the system due to this 
momentum kick are the atom laser output.  We assume that the total 
momentum kick is purely in the longitudinal direction, $x$.  We are 
also assuming that $1/t_{0} > \gamma_{4}$, where $\gamma_{4}$ is the 
linewidth of level $|4 \rangle$, so that the output level $|4 \rangle$ 
has a long lifetime.  Here, $t_{0}$ is estimated by
\begin{equation}
t_{0} = \frac{l_{x}}{2 \hbar k_{x}/m},
\label{t0}
\end{equation}
where $m$ is the mass of the atom, $l_{x}$ is the length of the 
light-atom cavity interaction region in the $x$ direction and $2 \hbar 
k_{x}$ is the size of the momentum kick.  To transfer the population 
out of the system, we consider a regime where the loss rate out of the 
interaction region, $1/t_{0}$ is larger than the Raman transition rate 
from level $|4 \rangle$ back to $|2 \rangle$.  In this regime 
effectively all atoms that transfer to level $|4 \rangle$ will exit 
the system.  This condition is given by
\begin{equation}
(N_{21}+1)
~r_{24} <  \frac{1}{t_{0}}, \label{Condition1}
\end{equation}
where $r_{24}$ is the Raman transition rate given in Eq.  (\ref{r24}).  
The factor $(N_{21}+1)$ is a Bose-enhancement factor for the backward 
process, due to the lasing mode having an occupation of $N_{21}$ 
bosons.  Here we have ignored the transitions into other higher energy 
modes of the lasing cavity, as in practice we find that only the 
ground state ($j=1$) mode has a significant non-zero population.

We require a coherent transfer of the population from level $|2 
\rangle$ to level $|4 \rangle$ for our atom laser output.  To ensure 
that the transfer is unitary we require negligible spontaneous 
emission from level $|3 \rangle$.  More specifically, we require that 
the rate at which atoms populate and then spontaneously emit from 
level $|3 \rangle$ is much less than the rate at which atoms leave the 
system coherently, $N_{21} r_{24}$.  This constraint is given by
\begin{equation}
\frac{\Omega_{1}^{2}}{\Omega_{1}^{2} + \Delta^{2}}
\frac{N_{21} \Gamma_{3}}{2} << N_{21} r_{24}.
\label{Condition2}
\end{equation} 
We have only considered in this expression the population of $|3 
\rangle$ due to atoms in level $|2 \rangle$.  The population due to 
level $|4 \rangle$ atoms is negligible compared to this, as the number 
of atoms in state $|4 \rangle$ is always small, (see Sec.  
\ref{sec:Equations}).  In the physical implementation discussed in 
Sec.  \ref{sec:Implementation} we evaluate Eq.  (\ref{r24}) for 
$r_{24}$ using values that satisfy constraints (\ref{Condition1}) and 
(\ref{Condition2}).

In the discussion above we assumed that the lasers impart a fixed 
momentum kick of magnitude $2 \hbar k_{x}$ to the atoms.  This 
assumption corresponds to modelling the lasers as plane waves.  Atoms 
undergo a Raman transition, absorbing a photon from one of the beams 
and emitting into the other.  In a plane wave model of the lasers, the 
final state of the atoms in the longitudinal direction will correspond 
to the initial wavefunction in the ground state of the lasing cavity 
combined with the momentum kick $2 \hbar k_{x}$.  In the transverse 
direction, the output state of the atoms remains the ground state mode 
of the hollow optical fibre.  This is achieved using output coupling 
lasers with a sufficiently narrow linewidth compared with the 
separation of the lasing cavity transverse energy levels.  By suitably 
tuning such Raman lasers to the output atomic level it is impossible 
for the atoms to excite into a higher transverse mode of the fiber, as 
the energy required to change internal atomic levels and excite the 
atoms to a higher transverse mode is higher than the available energy 
from the Raman photons.  For a fiber approximately $2 \mu \mbox{m}$ in 
diameter this requires the Raman lasers to have a linewidth of only a 
few kHz, which can be achieved by active stabilization.

In the longitudinal direction, the output atoms in state $|4 \rangle$ 
are no longer trapped.  The atoms couple to a continuum of momentum 
eigenstates.  Since the initial wavefunction is an energy eigenstate 
of the lasing cavity the atoms do not have a definite momentum, due to 
the position-momentum uncertainty principle.  Hence, when the atoms 
leave the trap they have a momentum distribution with some width 
$\hbar \Delta k$.  For our parameters this width is smaller than the 
longitudinal momentum imparted to the atoms, $2 \hbar k_{x}$.  Based 
on the time-energy uncertainty relation we expect the output linewidth 
to be narrower for slower output coupling.  We can obtain an upper 
bound for the output width by considering fast output coupling.  By 
fast we mean that the internal state of the atom is changed without 
time for the spatial wavefunction to evolve.  Then the spread in 
momentum due to the presence of a momentum distribution in the cavity 
is identical to the momentum spread one would obtain from a pulsed 
atom beam created by removing the walls of the cavity as suggested by 
Guzman {\it et al.} \cite{Guzman96}.

For the parameters considered in Sec.  \ref{sec:Implementation}, the 
longitudinal kick is of the order of $2 k_{x} \approx 1.3 \times 
10^{7}~ \mbox{m}^{-1}$, in comparison to the momentum spread of the 
atoms in the lasing mode of length $2 \mu \mbox{m}$ of $\Delta k 
\approx 10^{6} \mbox{m}^{-1}$.  A further factor which contributes to 
the final momentum spread of the atomic beam is the shape of the Raman 
laser beams.  We can estimate the size of this spread by considering a 
Gaussian laser beam focussed down to a waist of size $w_{0} = 2 \mu 
\mbox{m}$.  This corresponds to focussing onto an interaction region 
of the size of the laser cavity that we consider in Sec.  
\ref{sec:Implementation}.  The gaussian transverse wavevector spread 
has standard deviation $\sigma = 5 \times 10^{5} \mbox{m}^{-1}$.  Thus 
the lasers impart a range of kicks which is somewhat smaller than the 
spread in the momentum in the cavity, $\Delta k$.  We can further 
decrease this spread by increasing the beam waist.  This, however, 
increases the size of the interaction region, eventually invalidating 
inequality (\ref{Condition1}).  Nevertheless, even when it is invalid, 
some atoms will continue to couple out of the system.

\section{The Atom Laser Rate Equations} \label{sec:Equations}

The atom laser discussed here contains analogous elements to those 
found in the optical laser.  One of the differences is in the pumping 
process.  Here our pumping consists of the loading of the pump cavity.  
In an optical laser, pumping involves exciting atoms which then emit 
photons.  This difference is a consequence of the inability to create 
atoms in a manner analogous to creation of photons.  Instead, atoms 
must be transferred from different states to the lasing state.  
Another difference is that the output coupling of our atom laser also 
involves changing the state of the atoms.

One important characteristic of the optical laser that is observed in 
our atom laser model is the presence of a threshold condition.  This 
threshold condition occurs in an optical laser when the net 
amplification between the mirrors for a single photon circulating the 
cavity equals the loss at the mirrors.  Similarly for the atom laser 
threshold, we consider atoms injected into an otherwise empty system 
($N_{2j} = 0$).  The threshold condition occurs when the single atom 
input rate into the lasing cavity, $g_{1} r_{12}$, is just sufficient 
to dominate the loss rate, $r_{24}$.  We thus expect a threshold when 
$r_{12} = r_{24}/g_{1}$.  The threshold condition can be expressed in 
terms of the injection rate into the pump cavity, $r_{1}$, using the 
fact that for the empty system considered for the onset of threshold, 
$r_{1} \approx r_{12}$ at steady state.  This gives the threshold 
condition in terms of the rate of input into the pump cavity as $r_{1} 
\approx r_{24}/g_{1}$.

We now present rate equations for this atom laser scheme.  Similar 
equations have been investigated independently by Spreeuw {\it et al}.  
in the context of another atom laser scheme.  In the regime where all 
the atoms that transfer to level $|4 \rangle$ are lost from the 
system, we find that our rate equations reduce to a form equivalent to 
those of Spreeuw {\it et al.}, though the overlap factors $g_{j}$, and 
rates, $r_{12}$ and $r_{1}$ in our equations are different due to 
physical differences between the schemes.  Spreeuw {\it et al}.  do 
not consider a separate output atomic level for their scheme so they 
have no corresponding equation to our Eq.  (\ref{dn4dt}) for $N_{4}$.  
These rate equations allow us to investigate the number of atoms in 
the modes of the lasing cavity as a function of the pumping rate and 
of time, and to verify the threshold condition.  We are interested in 
the population of each of the combined atom and cavity states using 
realistic parameters for our laser model.  Using the notation 
presented earlier, rate equations for each of the atom laser levels 
are given by
\begin{mathletters}
\begin{eqnarray}
\frac{d N_{1}}{d t} &=&
r_{1} - \sum_{j} g_{j} (N_{2j} + 1) r_{12} N_{1}  
\nonumber\\
&&
\rule{27mm}{0mm} - (1-\sum_{j} g_{j}) r_{12} N_{1}
\label{dn1dt},\\
\frac{d N_{2j}}{d t} &=& g_{j} r_{12} N_{1} (N_{2j}+1) -
r_{24} N_{2j} 
\nonumber\\
&& \rule{27mm}{0mm} + ~G_{j} (N_{2j}+1) N_{4}
r_{24}, \label{dn2dt}\\
\frac{d N_{4}}{d t} &=& \sum_{j} N_{2j} r_{24}
-\sum_{j} G_{j} 
(N_{2j}+1) N_{4} r_{24} \nonumber\\
&& \rule{27mm}{0mm}  - N_{4} \frac{1}{t_{0}}. \label{dn4dt}
\end{eqnarray}

\end{mathletters}
Eq.  (\ref{dn1dt}) describes the pump level.  The first term gives the 
input rate into the pump level.  The second term corresponds to the 
transfer of atoms from the pump cavity to the lasing modes as 
described in Eq.  (\ref{TotRate}).  This term includes the bosonic 
enhancement of transitions into lasing states due to the presence of 
$N_{2j}$ bosons.  The final term gives the loss from level $|1 
\rangle$ into states which are not in the laser cavity.  Eq.  
(\ref{dn2dt}) describes the population of the various modes of the 
lasing cavity.  The first term corresponds to the Bose enhanced input 
into the levels of the lasing cavity from the pump.  The second term 
describes the loss from the lasing states into the output level, $|4 
\rangle$.  The final term describes a coupling of atoms from level $|4 
\rangle$ back into the lasing state.  Finally, Eq.  (\ref{dn4dt}) 
describes the population of the output level, $|4 \rangle$.  $N_{4}$ 
is the number of atoms in level $|4 \rangle$ which are still confined 
to the system.  The first term describes the gain in level $|4 
\rangle$ due to atoms transferring from the lasing states.  The last 
two terms describe losses out of level $|4 \rangle$ due to coupling 
back to the lasing cavity and out of the system respectively.

Level $|4 \rangle$ actually consists of a manifold of states $|4,j 
\rangle$, where state $|4,j \rangle$ corresponds to an atom that has 
made a Raman transition from the $j$th mode of the lasing cavity to 
the electronic level $|4 \rangle$.  Since the transformation is 
unitary each of these $|4,j \rangle$ states only couples back to its 
corresponding $|2,j \rangle$ state, where $|2,j\rangle$ describes an 
atom in the $j$th mode of the lasing cavity.  For simplicity of 
notation, we have not considered separate equations for the states 
$|4,j \rangle$.  Instead, we define a coupling constant, $G_{j}$, for 
transitions between $|4 \rangle$ and $|2,j \rangle$.  $G_{j} = N_{4j} 
/ N_{4}$ is the probability that an atom with an electronic level $|4 
\rangle$ is in the state $|4,j \rangle$.  We approximate $G_{j} 
\approx N_{2j}/N_{2}$ in the numerical calculations discussed below, 
however it is found that the results are insensitive to the value of 
$G_{j}$ as this back-coupling term is small in the regime defined by 
inequality (\ref{Condition1}).  This approximation for $G_{j}$ is 
based on the fact that the lasing atoms transfer to level $|4 \rangle$ 
at a rate which is independent of the lasing mode from which they 
come.  Since the lasing atoms are the only source for $|4 \rangle$, 
the number of atoms in each of the $|4,j \rangle$ states depends only 
on the number of atoms in the corresponding $|2,j \rangle$ state of 
the lasing cavity.

From Eqs.  (\ref{dn1dt}-\ref{dn4dt}), the steady state population of 
the lasing mode $N_{21}$ can be found.  The solution is somewhat 
complicated, though a simplified solution can be found for the case 
where we assume that all atoms that enter level $|4 \rangle$ are 
rapidly lost from the system, as required by inequality 
(\ref{Condition1}).  In this regime the steady state is given by
\begin{eqnarray}
N_{2j} &=& \frac{1}{2
g_{j}} \left[ \left(R_{j} - (1+\sum_{j' \neq j} g_{j'} 
N_{2j'} ) \right) +
\right. \nonumber\\
    && \left. \left( \left(R_{j} - (1+\sum_{j' \neq
j}
     g_{j'} N_{2j'})\right)^{2} + 
4 R_{j} g_{j} \right)^{\frac{1}{2}}
\right], \label{n2}
\end{eqnarray}
where the $R_{j}$ are dimensionless
pumping rate parameters, given by
\begin{equation}
R_{j}= \frac{r_{1}
g_{j}}{r_{24}}.
\end{equation}

The time dependent solutions of Eqs.  (\ref{dn1dt})-(\ref{dn4dt}) can 
be found numerically.  Fig.  \ref{dynamics} shows a logarithmic plot 
of the number of atoms in the lasing mode as a function of time, for 
an input pumping rate of $r_{1} = 1000 \mbox{ s}^{-1}$.  For the 
parameters of Sec.  \ref{sec:Implementation} it takes a time on the 
order of $10$ seconds for a large number $(\gg 1)$ of atoms to build 
up in the lasing mode.  After this time the number of atoms reaches a 
steady state value and remains constant.  The number of atoms 
populating the next highest energy state of the lasing cavity is also 
plotted in Fig.  \ref{dynamics}.  In this plot we see that due to gain 
competition and the Bose-enhancement of transitions into the highly 
populated ground state mode the steady state population of the next 
highest mode is negligible.  All higher modes are also found to have 
negligible population.  In the regime where the populations of all but 
the ground state mode are negligible, the steady state equations given 
in Eq.  (\ref{n2}) reduce to the form
\begin{equation}
N_{21} = \frac{1}{2 g_{1}} \left[ \left(R_{1} - 1
\right) + 
\sqrt{\left(R_{1} - 1\right)^{2} + 
4 R_{1} g_{1}}~ \right].
\label{n2simple}
\end{equation}
 
This result is analogous to the standard laser population equation 
\cite{Siegman} and equivalent to results of Spreeuw {\it et al.} 
\cite{Spreeuw95}.  In the limit of strong pumping $r_{1} \rightarrow 
\infty$ the number of atoms in the lasing mode, $N_{21}$, increases 
linearly with the pumping rate $R_{1}$, with a slope of $1/g_{1}$.  We 
assume numerical values for the transition rates, $r_{12}=0.1 \mbox{ 
s}^{-1}$, $r_{24}=0.125 \mbox{ s}^{-1}$ and for the wavefunction 
overlap, $g_{1}=0.00571$.  These parameters are justified in Sec.  
\ref{sec:Implementation} and correspond to a particular implementation 
of our scheme using hollow optical fibers.  A logarithmic plot of the 
number of atoms at steady state in the lasing cavity, $N_{21}$, as a 
function of the dimensionless pumping rate $R_{1}$ is given in Fig.  
\ref{threshold}.  The threshold pumping rate is $R_{1} = 1$, which 
corresponds to an input pumping rate, $r_{1} \approx 21.9 \mbox{ 
s}^{-1}$.

\section{Implementation} \label{sec:Implementation}

We propose here a possible implementation of this atom-laser model 
using hollow optical fibers.  A schematic diagram of this is shown in 
Fig.  \ref{fiber}.  Single mode hollow optical fibers, with holes of 
about $1.5 \mu \mbox{m}$ have been proposed for guiding atoms 
\cite{Marksteiner93,Savage91} and multimode fibers have already been 
demonstrated experimentally to guide atoms \cite{Ito95,Renn96,Ito96b}.  
The hollow fibre acts as a waveguide for atoms.  However in contrast 
with the optical case, the longitudinal atomic motion along the fiber 
decouples from the transverse motion, so there is a continuum of 
longitudinal plane wave modes which atoms can couple into.  A detailed 
development of the theory of hollow optical fiber waveguides is given 
by Marksteiner {\it et al.} \cite{Marksteiner93}.  Laser guiding of 
atoms using blue-detuned light has been observed by Ito {\it et al.} 
\cite{Ito96b}.  An enhancement of 20 times the ballistic flux was 
observed, indicating the guiding of atoms down a fiber.  The main 
mechanism which destroys the coherence in hollow-fiber atom optics 
experiments is spontaneous emission by atoms in the confining light 
field.  This is particularly a problem in those experiments that use 
red-detuned light as the atoms are continuously in the light field.  
However, for the blue detuned case the atoms only interact with the 
light field when they approach the inner wall of the hollow fiber.  
The effect of spontaneous emission while the atoms are in this light 
field can be reduced by increasing the detuning of the light field 
from the atomic resonance.  We assume confining light with blue 
detuning of $\Delta_{\mbox{max}} = 2 \pi \times 50 \mbox{THz}$.  We 
assume a typical linewidth of $\gamma = 2\pi \times 6 \mbox{MHz}$ for 
an (unspecified) upper level $| e \rangle$.  The spontaneous emission 
rate is given by the usual formula, $\Gamma_{\mbox{se}} = \gamma 
\Omega^2 / 4 \Delta^2$, where the Rabi frequency is determined by the 
required potential height.  As discussed by Hope and Savage 
\cite{Joe95} we assume that the minimum excited state population is 
limited by the maximum possible detuning, rather than by the available 
laser power.  Then the minimum possible spontaneous emission rate is 
given by Eq.  (14) of Hope and Savage \cite{Joe95} 
\begin{equation}
\Gamma_{\mbox{se,min}} = \frac{\gamma (k_b
T) }{\hbar \Delta_{\mbox{max}} }.
\label{minse}
\end{equation}
The required potential height has been expressed in terms of 
Boltzmann's constant $k_b$ and the atom temperature $T$.  With a 
temperature of $T=200 \mbox{nK}$ these parameters give a spontaneous 
emission rate of approximately $0.003 \mbox{s}^{-1}$.  Thus a typical 
atom must spend a time of order $300 \mbox{s}$ inside the confining 
light fields before a spontaneous emission event is likely.  Note that 
this time is an upper bound since we have assumed that the atom always 
experiences the maximum field.  Nevertheless it is possible to further 
reduce this spontaneous emission rate by a factor of almost $100$, to 
$6.0 \times 10^{-5} \mbox{s}^{-1}$, by using the Raman scheme of Hope 
and Savage \cite{Joe95} to create the potential.

The pump cavity is formed using light induced potentials from 
blue-detuned lasers shone transversely across the optical fiber.  
Because the pump cavity is long and narrow, the transverse mode energy 
level spacings are much larger than the longitudinal mode spacings.

The lasing cavity is also produced in the fiber by using two blue 
detuned lasers.  These are much closer together than in the pump 
cavity, producing a much larger energy level spacing.  Due to the 
large overlap of the lowest pump cavity modes with the ground state of 
the lasing cavity only it becomes significantly populated.  To 
maximize this overlap the input atoms must be pre-cooled to a few 
hundred nanokelvin so as to populate the lower energy states of the 
pump cavity.  Such temperatures can be achieved by evaporative 
cooling.

We calculate the average overlap for a pump cavity modeled as 
discussed in Sec.  \ref{sec:Input}.  The pump cavity is modeled as a 
square well with sides of length $100 \mu \mbox{m}$ in the 
longitudinal direction.  The lasing cavity is modeled as a three 
dimensional harmonic oscillator.  The oscillator frequencies 
$\omega_i$ are specified in terms of the ground state width, $d_{i}$ 
\begin{equation}
d_i = 2
\sqrt{ \frac{\hbar}{2m \omega_i} }.
\label{gswidth}
\end{equation}
In the longitudinal direction the width of the lasing cavity is $d_{i} 
= 2 \mu \mbox{m}$.  Transverse confinement for both cavities is 
modeled by oscillators in the transverse directions with widths of 
$d_{i} = 1.5 \mu \mbox{m}$.  The lasing cavity is displaced a distance 
$\bar{x}$ from the center of the pump cavity.  For definiteness, we 
consider a spontaneous emission kick from the $|1 \rangle \rightarrow 
|2 \rangle$ transition of magnitude $k_{0} = 10^{6} \mbox{ m}^{-1}$ 
which corresponds to an infrared transition.  Modeling a lasing cavity 
placed at the edge of the pump cavity, $\bar{x} = 48 \mu \mbox{m}$, we 
calculate the average overlap for a spontaneous emission kick of 
magnitude $10^{6} \mbox{ m}^{-1}$ to be $g_{1} = 0.00571$.  The 
overlaps, $g_{j}$ with the higher exited states ($j > 1$) are smaller, 
with the next greatest overlap, $g_{2} = 0.00362$ occuring with the 
state $n_{x} = 1$, $n_{y}=n_{z}=0$.  In the analysis leading to Fig.  
\ref{dynamics} we use these values, and consider the six lowest energy 
states of the lasing cavity.  All of these apart from the first are 
found to have negligible population in steady state because of their 
weaker overlap with the pump cavity modes, and due to the 
Bose-enhancement of transitions into already populated states.  In the 
rate equations we also use an atomic transition rate, $r_{12}$, given 
by the inverse lifetime of the initial atomic level, $|1 \rangle$.  
For definiteness we assume that $r_{12} = 0.1 \mbox{ s}^{-1}$, 
corresponding to $|1 \rangle$ being metastable.

The output coupling is achieved by shining two lasers diagonally 
across the lasing cavity.  This has the dual purpose of localizing the 
interaction region, as well as providing a total momentum kick, $2 
\hbar k_{x}$, directed along the longitudinal axis of the optical 
fiber (the transverse components cancel).  The atoms move out of the 
interaction region due to the momentum kick, thus forming the atom 
laser beam.  For definiteness we arrange a value of $2 k_{x} = 1.31 
\times 10^{7} \mbox{ m}^{-1}$.  This value could be achieved using 
lasers with wavelengths $\lambda = 480 \mbox{ nm}$ oriented at $60 
^{0}$ to the long axis of the fiber.  It is possible to produce light 
of this wavelength with a frequency doubled titanium-sapphire laser.  
Assuming a typical atomic mass $m = 10^{-26} \mbox{ kg}$ and size of 
the interaction region, $l_{x} = 2 \times 10^{-6} \mbox{ m}$, we find 
that the timescale on which atoms leave the system due to the momentum 
kick is given by Eq.  (\ref{t0}) to be $t_{0} = 1.5 \times 10^{-5} 
\mbox{ s}$.  Using this value for $t_{0}$, we find that constraint 
(\ref{Condition1}) can be fulfilled for lasing mode populations of 
$N_{21}$ up to $N_{21} \approx 10^{5}$ with large detunings, and 
relatively small Rabi frequencies.  If state $|3 \rangle$ has a 
linewidth of $\Gamma_{3} = 2 \pi \times 1.6 \mbox{ MHz}$ then both 
constraints (\ref{Condition1}) and (\ref{Condition2}) can be satisfied 
with the output lasers detuned by an amount, $\Delta = 2 \pi \times 
1.6 \mbox{ GHz}$ and Rabi frequencies, $\Omega_{1} = 2 \pi \times 50 
\mbox{ kHz}$ and $\Omega_{2} = 2 \pi \times 1.6 \mbox{ MHz}$.  With 
these values the single atom rate constant, $r_{24}$ is given by Eq.  
(\ref{r24}) to be $r_{24} = 0.125 \mbox{ s}^{-1}$.  The total rate at 
which atoms leave the system as the atom-laser beam is then given by 
$N_{21} r_{24}$.

We do not propose a particular atom in which our level scheme might be 
implemented.  One possibility is to use a metastable, rather than a 
hyperfine, level for the output state $|4 \rangle$.  Atoms with two 
accessible metastable levels, for the pump state $|1 \rangle$ and the 
output state $|4 \rangle$, include manganese.

\section{Conclusion}
\label{sec:Conclusion}

In this work we have analyzed an atom laser model and calculated 
values for the lasing population using an implementation of the scheme 
based on hollow optical fibers.  We have proceeded by analogy with the 
photon laser and have described a device that will produce a large 
number of atoms in a single atomic state and a method of coherently 
coupling a beam of atoms out of this device.  We found that the output 
energy spread is bounded by the momentum spread of the atoms in the 
lasing cavity.  Since the spread in momentum introduced by the Raman 
lasers is smaller than the lasing cavity spread, the Raman output 
coupling gives an output spectral density similar to that produced by 
dropping the walls of the cavity.  This is because an atom making the 
Raman transition changes to a non-trapped state where the atom is no 
longer affected by the confining potential.  An important advantage of 
the Raman scheme over switching the cavity walls is that atoms can be 
let out continuously.  Moreover, the Raman transition gives the atoms 
a longitudinal momentum kick.

Other atom laser schemes \cite{Wiseman95a,Wiseman95b} have found large 
output linewidths due to collisions.  This may increase the linewidth 
of our atom laser, though as our linewidth is already broad we do not 
expect this to be a limiting factor and have not considered this in 
our model.  A criticism that has recently been made of a number of 
atom laser schemes \cite{Guzman96} is the lack of consideration given 
to inelastic collisions.  Such collisions are used explicitly in a 
number of models \cite{Holland95,Wiseman95b,Guzman96} to provide the 
transitions to the lasing mode.  We assume large blue detuning to 
minimize the excited state population due to the confining light.  
This minimizes the interatomic interactions in the lasing cavity.  The 
Raman transition scheme of Hope and Savage \cite{Joe95} for generating 
mechanical potentials provides a method for further reducing 
spontaneous emission.

We have presented an atom laser model, discussing input and output 
coupling mechanisms as well as a possible implementation of our scheme 
in hollow optical fibers.  We found that for reasonable parameters we 
get a large build up of atoms in the lasing mode, above a threshold 
pumping rate, in a manner analogous to the threshold found in optical 
lasers.

\section*{Acknowledgement}

The authors would like to thank Dr.  J. Eschner for his many helpful 
discussions.

\section*{Appendix}
\label{sec:Appendix}

The overlap, $g_{ij}$ between an atom in the $i$th state of the pump 
cavity and the $j$th state of the lasing cavity is given by

\begin{eqnarray}
g_{ij} &=&  \int_{0}^{\pi} \frac{d\phi}{4 \pi}
~\int_{-\pi}^{\pi} 
d \theta~ \sin(\phi)  \times \nonumber\\
      & &
\rule{17mm}{0mm} \left| 
 \int d{\bf r} ~  \phi_{i}^{*}({\bf r})
\psi_{j}({\bf r}) ~\exp{\left[i 
{\bf k.r}\right]} \right|^{2}
\label{overlap2}
\end{eqnarray}

where the domain of the ${\bf k}$ space integrals is a spherical shell 
of radius $k_{0}$.  This accounts for the spontaneous emission of a 
photon with wavevector ${\bf k}$ of magnitude $k_{0}$.  For simplicity 
we assume that spontaneous emission is isotropic.  Other emission 
distributions are found to give similar overlap results for a 
particular $k_{0}$.  $\phi_{i}$ and $\psi_{j}$ are the wavefunctions 
of atoms in the $i$th and $j$th energy states of the pump and lasing 
cavity respectively.  The lasing states are modelled in this paper as 
the eigenstates of three dimensional harmonic oscillators.  The 
transverse mode structure of the pump cavity is likewise modelled as 
harmonic oscillators, while the longitudinal states are eigenstates of 
a square well.

\begin{figure}
\caption{Schematic diagram of atomic states and output coupling 
lasers.}
\label{EnergyLevels}
\end{figure}

\begin{figure}
\caption{Schematic diagram of a possible implementation of our atom 
laser model using a hollow optical fiber.  The lasing cavity is 
confined to a region of size $2 \mu \mbox{m}$.  The pump cavity is 
approximately $100 \mu \mbox{m}$ long.  The input coupling is by a 
partially transmitting atomic mirror for atoms in state $|1 \rangle$, 
indicated by the laser on the left of the figure.  Lasers are also 
used for the other end mirror of the pump cavity, and for the lasing 
cavity.  The two lasers $\omega_{1}$ and $\omega_{2}$ are the output 
coupling lasers.  They are localized to the lasing cavity, and provide 
a momentum kick along the longitudinal axis of the fiber as shown.}
\label{fiber}
\end{figure}

\begin{figure}
\caption{Plot of number of atoms in lasing mode $N_{21}$ (solid line) 
and in the next highest energy mode $N_{22}$ (dashed line) of the 
lasing cavity as a function of time, $t$ for an input pumping rate, 
$r_{1} = 1000 \mbox{ s}^{-1}$, average overlap, $g_{1} = 0.00571$ and 
$g_{2} = 0.00362$ and single atom output rate, $r_{24} = 0.125 \mbox{ 
s}^{-1}$.  The dimensionless pumping rate, $R_{1} = 45.7$.}
\label{dynamics}

\end{figure}

\begin{figure}
\caption{Plot of the steady state number of atoms in the lasing mode, 
$N_{21}$ as a function of the dimensionless pumping rate, $R_{1}$.  
Average overlap, $g_{1} = 0.00571$.  Threshold occurs at $R_{1}=1$.}
\label{threshold}
\end{figure}

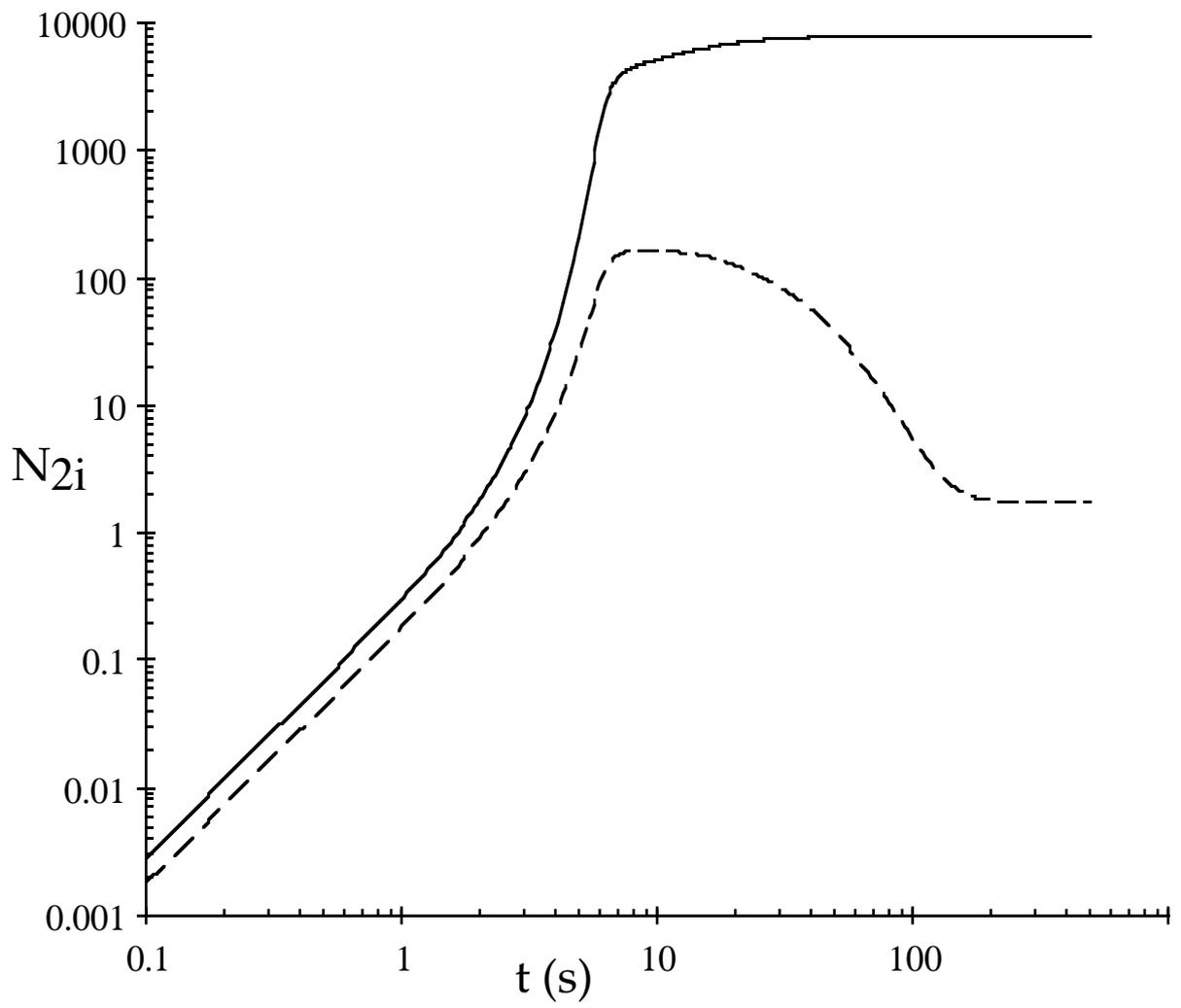
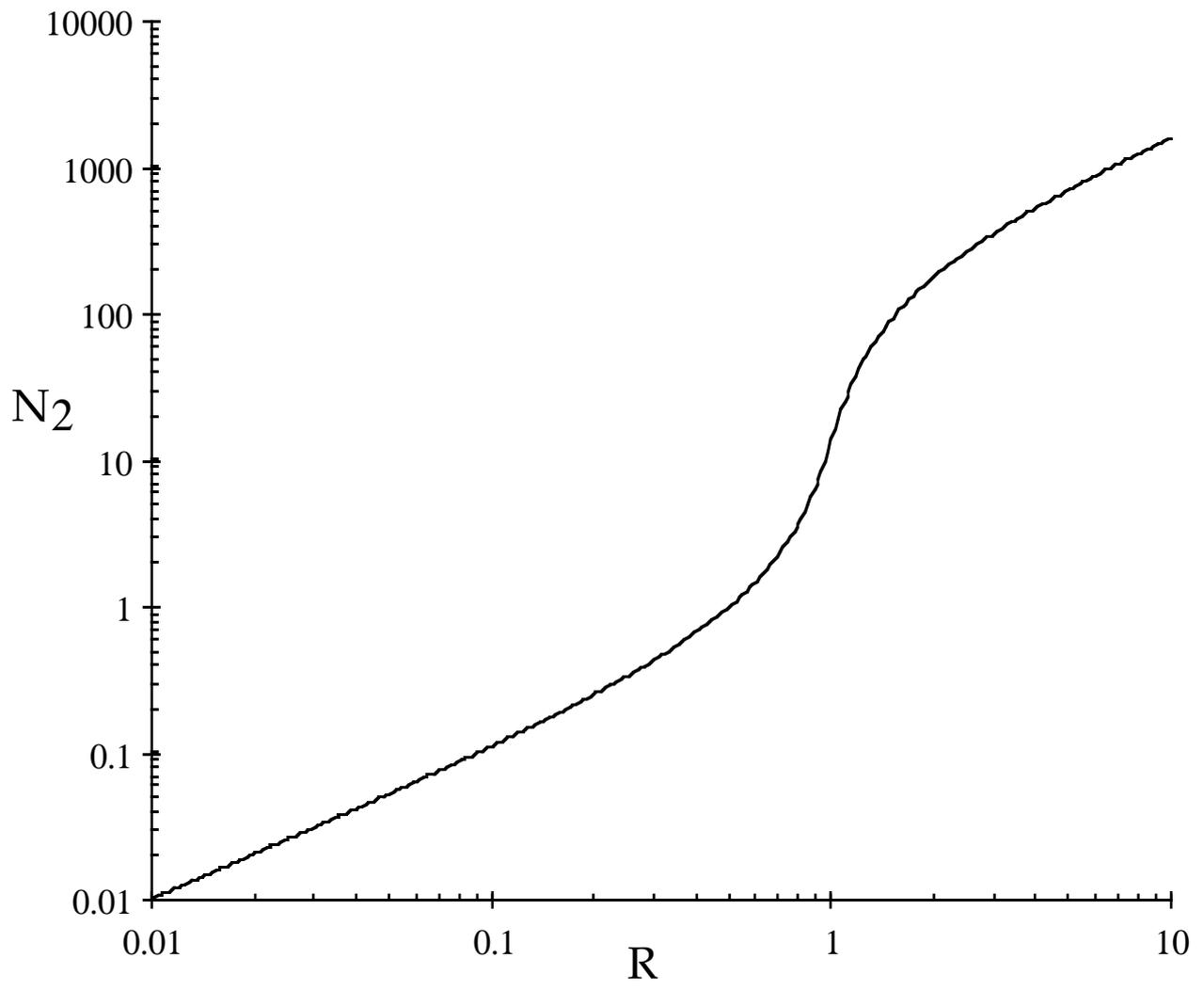
\end{document}